\documentclass[11pt,a4paper]{article}
\pdfoutput=1
\usepackage{jheppub}
\usepackage{graphicx}
\usepackage{multirow}
\usepackage{rotating}
\usepackage{amssymb,amsfonts,amsmath}

\newcommand{\bea}{\begin{eqnarray}}
\newcommand{\eea}{\end{eqnarray}}

\begin{document}

\preprint{
WUB/12-03
}

%Header ]]]
\title{Anisotropy tuning with the Wilson flow}
% Authors [[[

\author[a]{Szabolcs Bors\'anyi,}
\author[a,b]{Stephan D\"urr,}
\author[a,b,c]{Zolt\'an Fodor,}
\author[c]{S\'andor D. Katz,}
\author[a,b]{Stefan Krieg,}
\author[a]{Thorsten Kurth,}
\author[d]{Simon Mages,}
\author[d]{Andreas Sch\"afer,}
\author[a]{and K\'alm\'an K. Szab\'o}

\affiliation[a]{Bergische Universit\"at Wuppertal, Gaussstr.\,20, D-42119 Wuppertal, Germany.}
\affiliation[b]{J\"ulich Supercomputing Centre, Forschungszentrum J\"ulich, D-52425 J\"ulich, Germany.}
\affiliation[c]{Institute for Theoretical Physics, E\"otv\"os University, H-1117 Budapest, Hungary.}
\affiliation[d]{Universtit\"at Regensburg, Universit\"atsstr.\,31,
D-93053 Regensburg, Germany}

% Authors ]]]
\date{\today}
\abstract{ % [[[
We use the Wilson flow to define the gauge anisotropy at a given physical
scale. We demonstrate the use of the anisotropic flow by performing the
tuning of the bare gauge anisotropy in the tree-level Symanzik action
for several lattice spacings and target anisotropies.  We use this method to
tune the anisotropy parameters in full QCD, where we also exploit the diminishing
effect of a well chosen smearing on the renormalization of the fermion
anisotropy.
} % ]]]
\maketitle 
\section{Introduction} % [[[

Lattice QCD provided full, ab-initio answers for many questions of hadronic
physics e.g. the ligh hadron spectrum \cite{Durr:2008zz}. Nevertheless, there
are many questions, which are very difficult to answer. One characteristic
example is the ordering of the nucleon spectrum. In principle we have
techniques to understand ordering questions in spectrums (actually few groups'
results indicate a proper ordering for the nucleon states
\cite{Mathur:2003zf,Mahbub:2010rm}). The major issue in this example is to
obtain a good signal for the excited states with fine enough lattice spacing.
In order to minimize finite volume effects (by having large enough spatial
extensions) with fine enough spacings in the temporal direction one might take
larger lattice spacings in the spatial directions ($a_s$) than in the temporal
one ($a_t$). These asymmetric lattices are obtained by anisotropic bare
couplings.

Anisotropic lattice actions have a long history both for pure gauge theories
and for gauge plus fermionic systems.

In the quenched approximation anisotropic actions have been used to determine
the glueball spectrum \cite{Morningstar:1999rf}, to study heavy hybrids
\cite{Manke:1998qc,Harada:2002rf} and also for charmonium states
\cite{Chen:2000ej,Okamoto:2001jb,Klassen:1998ua}. The mostly advocated
technique to determine the lattice spacing asymmetry is based on the comparison
of spatial-spatial and spatial-temporal Wilson loops. A robust method for the
determination of the gauge anisotropy is even more important in dynamical
simulations, where less statistics are available for tuning.

For our purposes this dynamical case is the more important one. Dynamical
simulations have been done for the first time by the CP-PACS collaboration
using the Iwasaki gauge and clover improved fermion action \cite{Umeda:2003pj}.
The TrinLat collaboration used a Symanzik-improved gauge action and a Wilson
fermion action with a Hamber-Wu term \cite{Morrin:2006tf}.  Edwards et al. used
a clover-improved fermionic action with stout-link smearing (in the spatial
directions only) and a Symanzik-improved gauge action \cite{Edwards:2008ja}.
The latter is probably the most extensively used action today and led to
several interesting results (e.g. light hadron \cite{Lin:2008pr} or excited and
exotic charmonium \cite{Liu:2012ze} spectroscopy). Another important
application is to study spectral functions at non-vanishing temperatures
\cite{Aarts:2007pk}. In this case the many points of the meson correlators in
the temporal direction helps to determine the spectral function when one uses
the Maximum Entropy Method (MEM).

In the present paper we suggest an anisotropic action, which we plan to use in
our non-vanishing temperature studies. It is very similar to the isotropic
action, which we used in the the Budapest-Marseille-Wupperal collaboration for
determining e.g. the light hadron spectrum \cite{Durr:2008zz}, quark masses
\cite{Durr:2010vn,Durr:2010aw} or the transition temperature
\cite{Borsanyi:2012xx}. After defining the action we show how to set the
anisotropy parameters. In the interacting discretized theory the anisotropy
parameters in the action (bare anisotropies) differ from the observed
asymmetry, which is usually determined through the comparison of time and
space-like correlation lengths. Ignoring the radiative corrections to the
anisotropy parameters will introduce discretization errors that depend on the
logarithm of the lattice spacing, making a continuum extrapolation practically
impossible. Instead of the most popular choice, using Wilson loops to determine
the $a_s/a_t$ asymmetry in the gauge sector, we apply the Wilson flow. In the
fermionic sector the mass ratios of the pseudoscalars are used.

% Introduction ]]]

\section{\label{sec:aniflow}Wilson flow on anisotropic lattices} %[[[

In the continuum the Yang-Mills or Wilson flow is the solution of the differential equation
\bea
\frac{dA_\mu}{d\tau}= D_\nu F_{\nu\mu}
\eea
for the gauge field $A_\mu(x,\tau)$ supplemented with an initial condition at
$\tau=0$. The variable $\tau$ parametrizing the flow has a dimension of length squared.
Expectation values of operators along the Wilson flow have been a subject
to recent stuides in the SU(N) theory as well as in full QCD.
It was shown \cite{Luscher:2011bx} that for any $\tau>0$ time the field defined by the flow is
renormalized, no UV divergences appear to any order in perturbation theory.

On the lattice the flow was investigated by Luscher \cite{Luscher:2009eq}
primarily to study the behaviour of gauge field updating algorithms. It was
considered earlier by Narayanan and Neuberger \cite{Narayanan:2006rf} in a
different context. The discretization of the flow equation gives
\bea
\label{eq:flow_du}
\frac{dU_\mu}{d\tau}= X_\mu(U) U_\mu
\eea
where $X_\mu$ is the generator of the stout smearing
transformation \cite{Morningstar:2003gk}:
\bea
\label{eq:flow_x}
X_\mu(x,\tau)= \mathcal{P}_A\left[ \sum_{\pm\nu\ne \mu} \rho_{\mu\nu}U_\nu(x,\tau) U_\mu(x+\nu,\tau) U^\dagger_\nu(x+\mu,\tau)U^\dagger_\mu(x,\tau) \right],
\eea
with $\mathcal{P}_A$ operator projecting onto traceless, anti-hermitian
matrices, and in this case the smearing parameters are $\rho_{\mu\nu}=1$. The
flow variable $\tau$ is made dimensionless using the second power of the
lattice spacing (ie. $\tau/a^2 \to \tau$ when discretizing the flow). The
simplest way of implementing the Wilson flow is to make successive stout
smearing steps on a gauge field configuration with a small enough smearing
parameter.  Note, that there exist sophisticated integrators for the flow
like the third-order method introduced in \cite{Luscher:2010iy}. 

%TODO
%We have
%published a reference implementation with a support for anisotropic lattices
%along with Reference \cite{Borsanyi:2012zs}.

It was also realized \cite{Luscher:2010iy}, that the Wilson flow provides a
length scale, called $\sqrt{t_0}$, which can be used to set the scale in
lattice simulations. In \cite{Borsanyi:2012zs} we derived another scale from
the Wilson flow, $w_0$. It is defined by the following equation:
\bea
\label{eq:w0}
\left[\tau\frac{d}{d\tau} \tau^2 \langle E(\tau) \rangle\right]_{\tau=w_0^2} = 0.3,
\eea
where $\langle E(\tau)\rangle$ is the quantum expectation value of the Yang-Mills action density
\bea
E(\tau)= \frac{1}{4}\sum_x F_{\mu\nu}^2(x,\tau).
\label{eq:E}
\eea
This new scale was shown to be advantageous by many means:
it can be measured with high precision on the lattice, its
definition is free from fitting and extrapolation, it has only small quark mass
dependence and it is not sensitive to the details of the lattice discretization. Since
$w_0$ is not directly measurable in experiments, in \cite{Borsanyi:2012zs} we
also calculated $w_0$ in phyiscal units using previous lattice QCD data and obtained
$w_0= 0.1755(18)(04)$~fm.

In the following let us consider an anisotropic lattice, ie. let the ratio of
lattice spacings in the spatial and temporal directions be different.
The observed anisotropy of the gauge configurations we denote by
$\xi_g=a_s/a_t$.  Discretizing the continuum flow equation on this lattice
yields the same as Equations (\ref{eq:flow_du}) and (\ref{eq:flow_x}).  The
flow variable is now made dimensionless by the spatial lattice spacing (ie.
$\tau/a_s^2 \to \tau$). The difference to the isotropic case is, that the
smearing coefficents have to be chosen as $\rho_{i4}=\xi_g^2$ and
$\rho_{ij}=\rho_{4i}=1$ in order to obtain the correct flow equation in the
continuum limit.

In a non-interacting theory there is only one anisotropy parameter: $\xi_g$,
which also enters into the lattice action.  When an interacting theory is
discretized on an anisotropic lattice, the action is written in terms of bare
anisotropy parameters (eg. $\xi_g^{(0)}$ bare gauge anisotropy) and
unrenormalized fields. The bare parameters describe the theory on the scale of
the lattice spacing. $\xi_g$ will then be termed as renormalized gauge
anisotropy, it can be measured from gauge observables on the physical scale.
Since for any time $\tau>0$ the gauge field along the Wilson flow is already
renormalized, we expect, that the anisotropy parameter in the Wilson flow is
the renormalized gauge anisotropy $\xi_g$.

% Wflow on aniso ]]]

\section{\label{sec:xig}Scale and gauge anisotropy from the Wilson flow} %[[[

Similarly to the isotropic case the Wilson flow offers a convenient scale
setting procedure on the anisotropic lattice, too. Additionally it also offers
a way to determine the renormalized gauge anisotropy. We write the spatial
and temporal contribution of the action density in Equation (\ref{eq:E}) separately
as
\begin{eqnarray}
\label{eq:Es}
E_{ss}(\tau)&=&\frac{1}{4}\sum_{x,i\ne j} F_{ij}^2(x,\tau)\,,\\
E_{st}(\tau)&=&\frac{1}{2}\sum_{x,i}F_{i4}^2(x,\tau)\,.
\label{eq:Et}
\end{eqnarray}
In physical units the expectation values of these two parts are equal.
Since for any $\tau>0$ these operators are renormalized,
they offer a definition for the renormalized gauge anisotropy: $\xi_g^2=a_s^2/a_t^2$
can be defined as 
the ratio of the
field strength tensors in lattice units at some point along the flow 
$\langle
a_s^4E_{ss}(\tau)\rangle/\langle a_s^2 a_t^2E_{st}(\tau) \rangle$.
From now on we work in lattice units, ie. $a_s^4E_{ss} \to E_{ss}$,
$a_s^2 a_t^2E_{st}\to E_{st}$, $\tau/a_s^2 \to \tau$ and $w_0/a_s \to w_0$.
Instead of working with the field strength tensors directly, we will consider
the derivative of these tensors along the flow, ie. instead using
$\langle E_{ss}\rangle/\langle E_{st}\rangle$
we define the ratio
\begin{equation}
R_E=\left. \left[\tau\frac{d}{d\tau} \tau^2 \langle E_{ss}(\tau) \rangle\right]_{\tau=w_0^2}\middle/\left[\tau\frac{d}{d\tau} \tau^2 \langle E_{st}(\tau) \rangle\right]_{\tau=w_0^2}\right..
\label{eq:RE}
\end{equation}

Let us now turn to our definition of the $w_0$-scale and the renormalized anisotropy using the Wilson-flow. We use the spatial part of Equation (\ref{eq:w0})
to define the $w_0$-scale
\bea
\label{eq:w0s}
\left[\tau\frac{d}{d\tau} \tau^2 \langle E_{ss}(\tau) \rangle\right]_{\tau=w_0^2} &=& 0.15\,,
\eea
and we define the renormalized gauge anisotropy through the $R_E$ ratio:
\begin{equation}
\xi_g^2 = R_E,
\label{eq:xig}
\end{equation}
analogously to $\xi_g^2= \langle E_{ss} \rangle/\langle E_{st} \rangle$.
The calculation of $R_E$ itself requires the knowledge of the anisotropy
$\xi_g$ which enters into the discretized flow equation (\ref{eq:flow_du}).
Therefore Equations (\ref{eq:w0s}) and (\ref{eq:xig})
become a set of coupled equations for the unknown lattice anisotropy and
$w_0$-scale. 

To solve these equations, and find $\xi_g$ for an ensemble of
gauge configurations with unknown anisotropy one evaluates the flow with
various anisotropy paramteres $\rho_{i4}=\xi_w^2$ and $\rho_{ij}=\rho_{4i}=1$
in Equation (\ref{eq:flow_x}).
For each $\xi_w$ one first locates the flow time, where
Equation (\ref{eq:w0s}) holds and then calculates the ratio in Equation (\ref{eq:RE}).
Then one searches for the solution of the equation
\begin{equation}
R_E(\xi_w)/\xi_w^2=1\,,
\label{eq:REwithxiw}
\end{equation}
which provides the gauge anisotropy $\xi_g=\xi_w$.

\begin{figure}
\centerline{\includegraphics{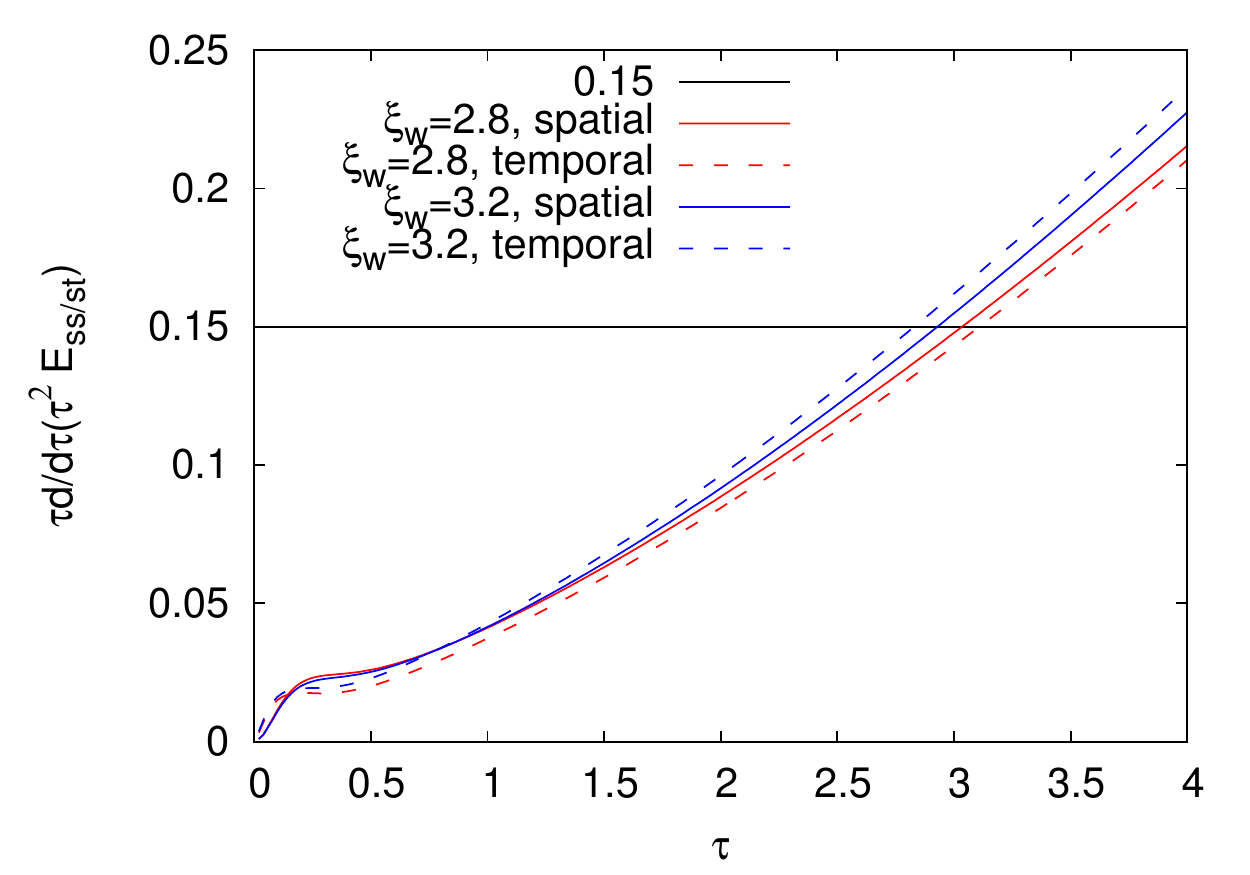}}
\caption{\label{fig:wflow} 
The derivative of the action density along the anisotropic Wilson-flow
at two different anisotropy parameters,
$\xi_w=2.8$ and $3.2$. The renormalized anisotropy is between these
two values. The temporal action density has been multiplied by $\xi_w^2$,
so that the spatial and temporal curves are similar in magnitude.
Notice that the temporal and spatial parts switch order between the two
flow anisotropies. In Figure \ref{fig:xig} we will define the renormalized
gauge anizotropy as $\xi_w$ at which the two curves coincide at $\tau=w_0^2$.
(Parameters: $\beta=6.1$, plaquette action with a bare
anisotropy of $\xi_g^{(0)}=2.46$.)
}
\end{figure}

The procedure of measuring the gauge anisotropy is illustrated on a quenched
ensemble generated with plaquette action with bare gauge anisotropy parameter
$\xi_g^{(0)}=2.46$ (see the Appendix for the definition of the action). 
Here we use a lattice size of $28^3\times 84$. Figure \ref{fig:wflow} shows
$\tau d \tau^2 \langle E_{ss} \rangle/d \tau$ and $\tau d\tau^2 \langle E_{st}\rangle/d\tau$ along the Wilson
flow two different Wilson flow anisotropies. In the figure $\langle E_{st} \rangle$ is rescaled
by $\xi_w^2$. $\langle E_{st} \rangle$ and $\langle E_{ss} \rangle$ are in the same units only if $\xi_g=\xi_w$,
but at this point we do not yet know $\xi_g$.

\begin{figure}[t]
\centerline{\includegraphics{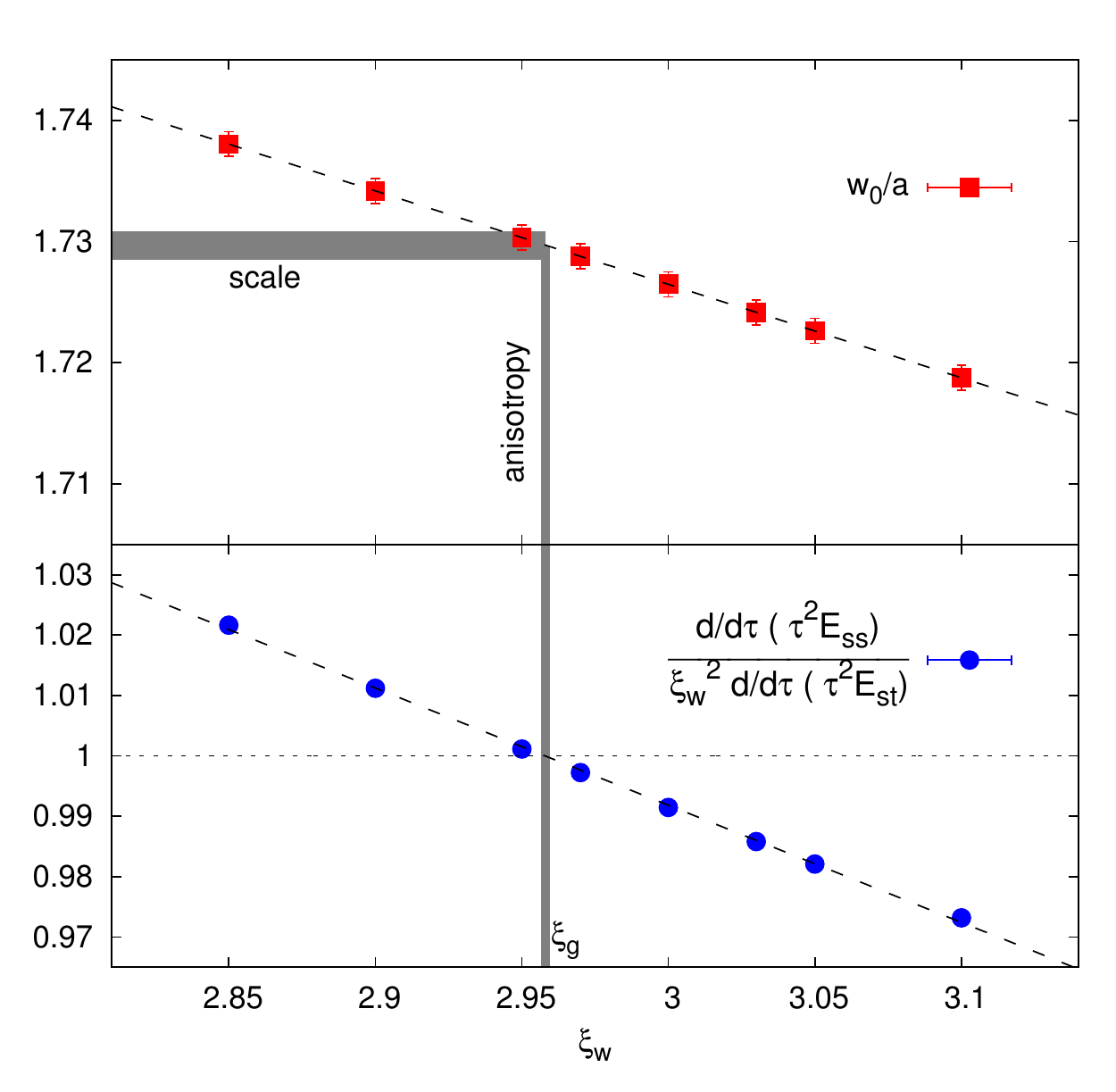}}
\caption{\label{fig:xig} The determination of the gauge anisotropy from
the Wilson flow. The same set of gauge configurations have been analyzed
with various $\xi_w$ parameters. For each $\xi_w$ the (spatial) $w_0$ scale
was determined through Equation (\ref{eq:w0s}). From each flow, the ratio $R_E$
in Equation (\ref{eq:RE}) was evaluated at the respective scale $\tau=w_0^2$.
We define the (renormalized) gauge anisotropy $\xi_g$ as the $\xi_w$ flow
anisotropy parameter at which the ratio ($R_E$) of the field strength
derivatives is equal to $\xi_w^2$. In this plot we divided this ratio by the
square of the actually used $\xi_w$, so that the fulfilment of the defining
condition is marked by a unit value of this combination. We then interpolate
$w_0/a$ to the newly defined $\xi_w=\xi_g$ point and use it as a
scale setting observable.
}
\end{figure}

Figure \ref{fig:xig} explains the determination of $\xi_g$.  We integrated the
Wilson flow on the same ensemble several times, each time with a different
$\xi_w$ parameter.  The left hand side of Equation (\ref{eq:REwithxiw}) is 
plotted as a function of $\xi_w$. In each case the ratio was
defined at the respective $\tau=w_0^2$ scale of the corresponding flow.
The self-consistency condition (\ref{eq:REwithxiw}) translates in the lower
plot to the crossing of the dotted line at 1.0.
Where this occurs defines the actual gauge anisotropy $\xi_g=\xi_w$. The
interpolation between the analyzed flows brings in no difficulty, the
dependence on $\xi_w$ is remarkably linear.  For this particular ensemble we
find $\xi_g=2.958(3)$ for the gauge anisotropy and $w_0/a=1.730(1)$ for the
$w_0$-scale in spatial lattice units. This converts to a lattice spacing of
$a_s\approx 0.102$~fm.

Let us close this section with a remark.  We use the derivatives of the field
strength tensors in Equation (\ref{eq:RE}) and in Equation (\ref{eq:w0s}). We
found that although the plain ratio $\langle E_{ss}\rangle/\langle E_{st} \rangle$ could
be an equally correct measure of anisotropy, its scaling features are
sub-optimal, even more so as it was in the case of the scale setting.  We
checked on several examples that as $\tau\to\infty$ the ratio  $\langle
E_{ss}(\tau)\rangle/\langle E_{st}(\tau)\rangle$ does indeed converge to
$R_E(\tau)$, but the $R_E(\tau)$ saturates much faster with growing flow time
$\tau$ as the simpler ratio. In fact, it is not necessary to define the
anisotropy through the $\tau\to\infty$ asymptotic behaviour, one may define it
at any fixed physical scale, like $\tau=w_0$.

\section{\label{sec:test}Comparison to Klassen's method}%[[[

Our next task is to compare the continuum scaling behaviour of our new gauge
anisotropy determination with an existing method. 
In the literature the gauge anisotropy is usually calculated
from ratios of Wilson loops:
\bea
R_{ss}(x,y)= \frac{W_{ss}(x,y)}{W_{ss}(x+1,y)},\\
R_{st}(x,t)= \frac{W_{st}(x,t)}{W_{st}(x+1,t)}
\label{eq:Rst}
\eea
the so-called Klassen-ratios \cite{Klassen:1998ua}. The anisotropy is obtained
from requiring, that for a given $x$ and $y$ the two ratios are equal at
$t=y\cdot\xi_g$:
\bea
\label{eq:klassen}
R_{ss}(x,y)= R_{st}(x,y\cdot\xi_g).
\eea
In practice one averages the so obtained $\xi_g$'s for different $x$ and $y$
values, for which the ratios are already in the asymptotic regime. The major
problem with this method is that one has to go beyond a few lattice spacings both
for $x$ and $y$ to avoid excited state contributions. However, measuring these
ratios becomes notoriously difficult as the size of the Wilson-loop is
increased. A further difficulty arises for non-integer $\xi_g$, since
the condition in Equation (\ref{eq:klassen}) requires the interpolation
of the ratio (\ref{eq:Rst}). Since the ratio exists for integer $t$ values only
this interpolation is always ambiguous.

For the comparison we again use the plaquette action, where Klassen's tuned
bare anisotropies are known to reproduce $\xi_g\approx 3$.  In addition to the
data in Figure \ref{fig:xig} we generated three more ensembles in the $\beta$
range [5.8,6.2].  We calculated $\xi_K$ from Equation
(\ref{eq:klassen}).  The determination of $\xi_K$ includes a simultaneous
polynomial fit of both ratios ($R_{ss}(x,y)$ and $R_{st}(x,\xi_K y)$) for each
$x$, with the $y>x$ restriction. $\xi_K$ was defined by the minimum of the
global $\chi^2$ of the fit for any given $x$. As a last step, we performed an
asymptotic fit in the Wilson loop size parameter $x$.

For our Wilson-flow based anisotropy we integrated the Wilson flow several
times with various $\xi_w$ parameters and determined $\xi_g$. The solution of
Equation (\ref{eq:REwithxiw}) requires the interpolation of the data obtained
with various flow parameters.  In contrast to the case of the Klassen ratios,
here the interpolation can be made arbitrarily precise by increasing the number
of $\xi_w$ parameters at which the flow is integrated.  The point in using the
Wilson-flow (or continuous smearing) was in fact to select a scale of interest
in both directions independently, and without being restricted to integer
multiples of the lattice spacing. The absence of delicate fits enables a great
level of automatization. 

\begin{table}[h]
\begin{center}
\begin{tabular}{c||c|c|c|c}
$\beta$&5.8& 6.0& 6.1& 6.2\\
$\xi_g^{(0)}$&2.38&2.44&2.46&2.49\\
lattice&$24^3\times72$&$24^3\times72$&$24^3\times48$&$32^3\times 96$\\
\hline
$\xi_g$&2.917(2)&2.942(5)&2.958(3)&2.979(1)\\
$\xi_K$&2.91(4)&3.01(2)&2.99(3)&3.01(2)\\
$w_0$&1.006(1)&1.465(1)&1.730(1)&2.015(3)\\
\end{tabular}
\end{center}
\caption{\label{tab:plaq}
Ensembles used to compare our $\xi_g$ to $\xi_K$.
}
\end{table}

\begin{figure}
\begin{center}
\includegraphics{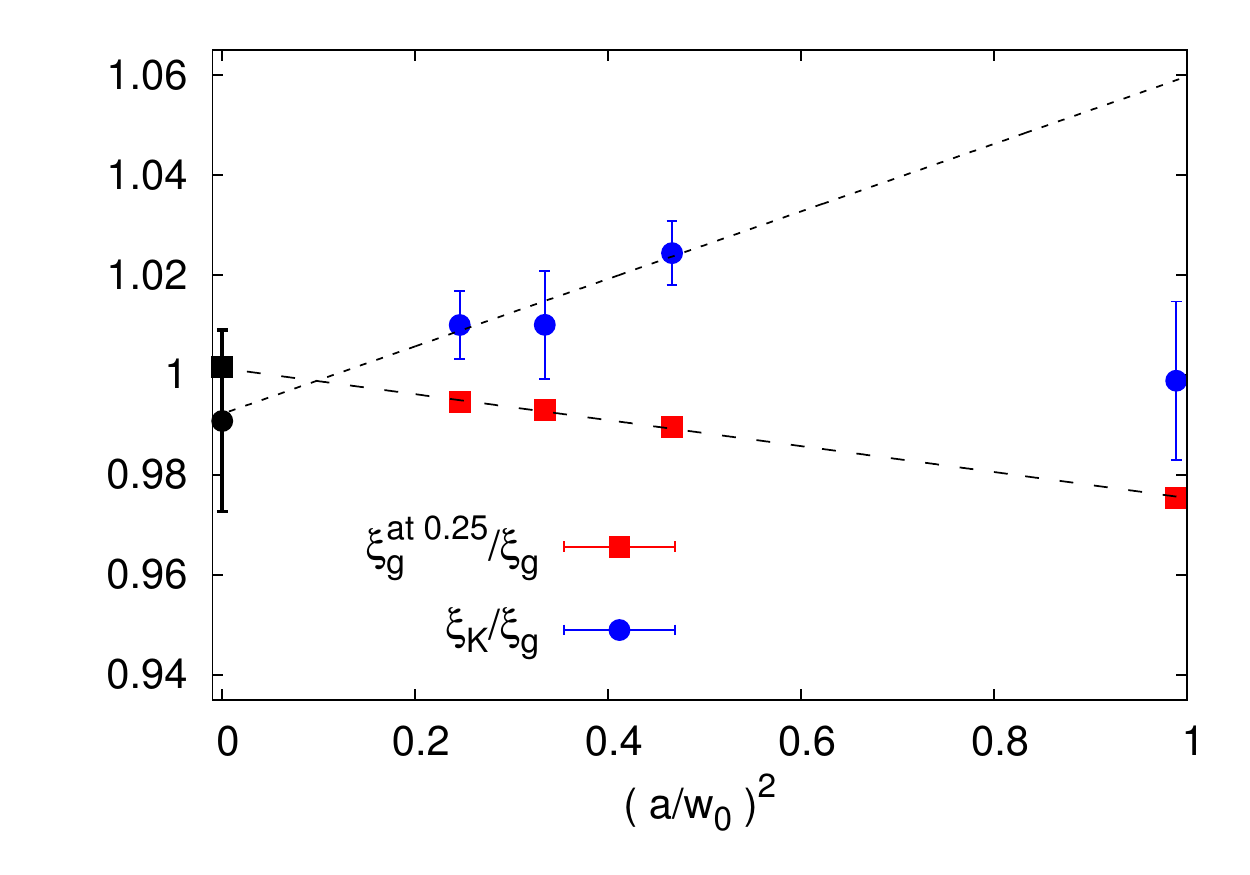}
\caption{\label{fig:xicmp}
The ratio of various gauge anisotropy definitions extrapolate to one in
the continuum. Blue circles: the anisotropy from the Klassen ratios divided
by our definition. Red squares: an other possible definion with 0.25
on the right hand side of Equation (\ref{eq:w0s}), again plotted relative to our $\xi_g$.
}
\end{center}
\end{figure}

Table \ref{tab:plaq} and Figure \ref{fig:xicmp} summarizes the result of the comparison.
We see deviations between the two anisotropy definitions on the percent
level. The errors on $\xi_K$ are larger than on $\xi_g$, and some of
the systematics are not controlled to our satisfaction.  Different definitions
for the gauge anisotropy do not need to agree for any one ensemble, but in the
continuum limit.  Figure \ref{fig:xicmp} shows the ratio of $\xi_K/\xi_g$ as a
funciton of $a^2$. If the coarsest lattice is not included in the extrapolation,
the continuum limits are compatible.

Figure \ref{fig:xicmp} shows an other comparison, too. A clear point of
ambiguity in our scheme is the choice of the scale at which $R_E$ and through
that $\xi_g$ is defined. Equation (\ref{eq:w0s}) defines the $w_0$ scale for
anisotropic lattices, but instead of the constant 0.15 any other positive number
could stand there. This constant tunes whether the anisotropy is renormalized
at $0.1755$~fm or at a different length scale.
For these ensembles we determined the gauge anisotropy with 0.25 on the right
hand side of the scale definition, resulting in a $\approx 21\%$ increase
in the renormalization scale. In Fig~\ref{fig:xicmp} we show in red the
ratio of this alternative result to our original definition.  We find that the
continuum extrapolation of the ratio very strictly follows an $a^2$ behaviour
and hits one in the continuum to per mill accuracy.

In principle, any constant in Equation (\ref{eq:w0s}) results in a valid definition,
although the choice of this constant may influnce the range of available
lattice spacings: on very fine lattices the flow time will grow unpractically
long, and on the other side, the coarse lattices may be outside of the scaling
regime. We made our choice to select the phenomenologically relevant range of
applicability.

A crucial ingredient in our definition is the use of a physical scale. This is
in contrast to more conventional schemes, where one calculates the anisotropy
at a scale fixed in lattice units. Then one extrapolates the scale to the far
infrared, as much as possible. It is technically feasible to fix the scale of
anisotropy renormalization in lattice units, say $\tau=9a^4$. We found, however,
that although such a scheme deviates form our discussed definition on the
percent level only, the discretization errors do not shrink as $\sim a^2$.
The discretization ambiguities in our final definition, which is based on the
$w_0$ scale, do indeed scale as $\sim a^2$ as we present in the next Section.

\section{\label{sec:universal}Universality of the anisotropic flow}%[[[

So far we have only considered the isotropic Wilson flow, discretized
on an anisotropic lattice. Here we discuss the opposite situation,
where the Wilson flow is anisotropic in the physical sense.

Thinking of the Wilson-flow as an UV-filter the flow equation's 
anisotropy parameter $\xi_w$ sets the ratio of the smearing radii. 
In an isotropic setting at flow time $\tau$ lattice modes
outside of a four-sphere in momentum space of a radius ($\sim{\tau}^{-1/2}$)
are suppressed. The anisotropic Wilson flow suppresses modes with momenta
outside of a four-ellypsoid.  If $\xi_w$ is set to the gauge anisotropy defined
at the scale $\sim\sqrt{\tau}$, i.e. $\xi_w=\xi_g$, then the radii of this
ellypsoid will be equal in physical units, and the flow will be isotropic in
the physical sense. 
Only if the flow is isotropic in this physical sense, can one assume that
the temporal and spatial parts of the action densities (and their derivatives)
are equal in physical units, i.e. $E_{ss}/E_{st}=\xi_g^2$ or $R_E=\xi_g^2$.

Setting $\xi_w$ independently of the actual anisotropy $\xi_g$ one can easily
work out the tree-level formulas for the action densities at linear order in
$\xi_w^2-1$
\begin{eqnarray}
\label{eq:Esslatpert}
E_{ss}^{\rm tree~level}(\tau)&=&\frac{g^2(N_c^2-1)}{256\pi^2\tau^2}\left[3-(\xi_w^2/\xi_g^2-1)\right]\\
\label{eq:Estlatpert}
E_{st}^{\rm tree~level}(\tau)&=&\frac{1}{\xi_g^2}\frac{g^2(N_c^2-1)}{256\pi^2\tau^2}\left[3-2(\xi_w^2/\xi_g^2-1)\right]
\end{eqnarray}
where $N_c$ is the number of colors. One may use running coupling constant $g$ 
evalulated at $\mu=\sqrt{8\tau}$ scale \cite{Luscher:2010iy}, though the
consistent treatment of the running coupling would require higher orders in the
$E$'s perturbative expanison.  

We do not expect our lattice data to be in the pertrubative regime where
these formulae apply. We quoted these continuum results to emphasize that
flows with a non-trivial anisotropy $\xi_w$ can be studied independently of the
anisotropy of the lattice. Actually the simplest way to study the
anisotropic flow is to use isotropic configurations, where $\xi_g=1$ is
granted.

Encouraged by the finiteness of the perturbative formulae, which is valid
for any $\xi_w/\xi_g$, we calculate the ratio $R_E/\xi_w^2$ in the
non-pertbative regime with simulations of the SU(3) theory. Our hypothesis
is that $R_E/\xi_w^2$ has a well defined continuum limit for any $\xi_w/\xi_g$.

\begin{figure}
\begin{center}
\includegraphics{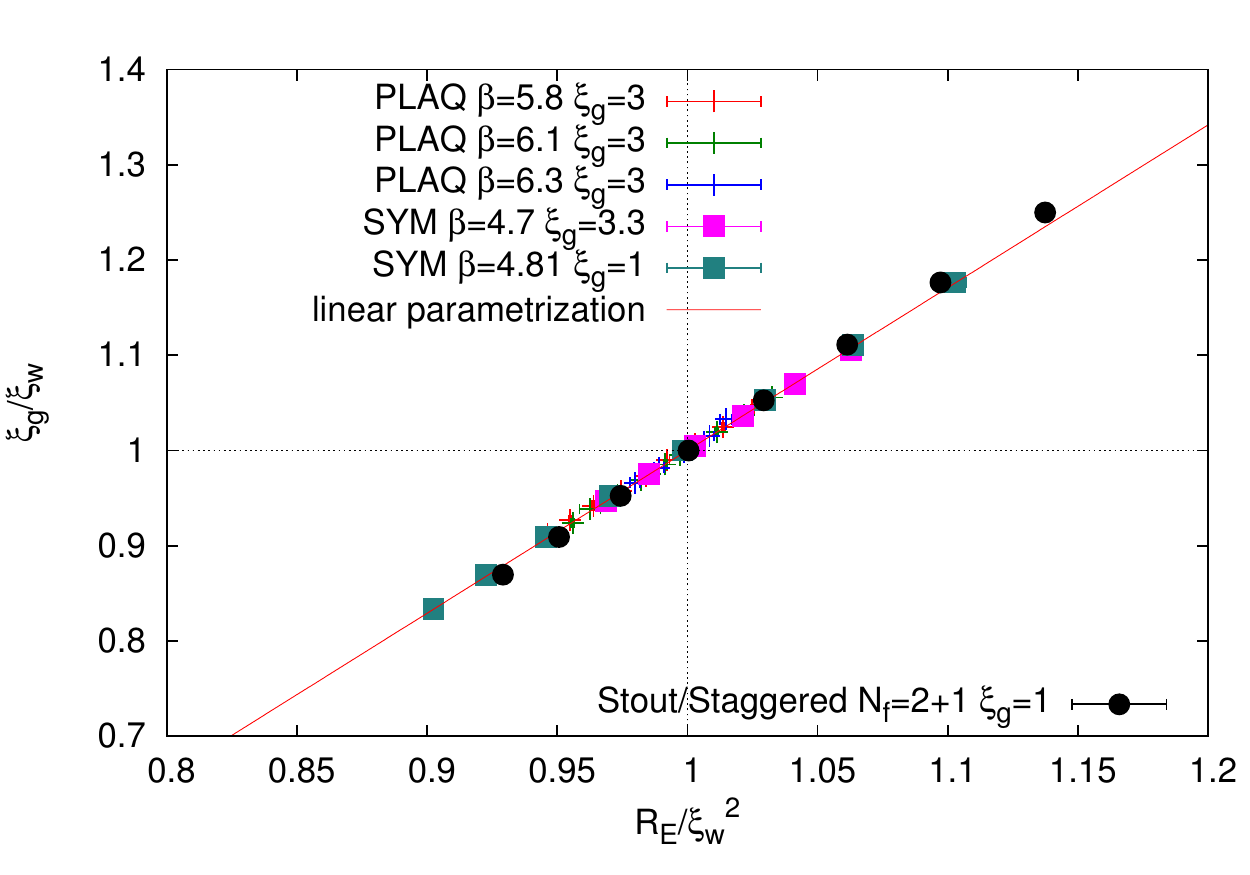}
\caption{\label{fig:universal}
The ratio of anisotropies $\xi_g/\xi_f$ and $R_E/\xi_w^2$ are connected
by a universal function. The staggered data set indicates that the effect
of unquenching is rather mild for this observable.
}
\end{center}
\end{figure}

To collect numerical evidence on our hypothesis we calculated the $R_E$ ratio
at several $\xi_w$ parameters.  This enabled us to know $\xi_g$ as well.
In Figure \ref{fig:universal}  we plot $\xi_g/\xi_w$ as a function of $R_E/\xi_w^2$.
If the flow is isotropic in physical units, both ratios are equal to one.
The curve, (which is very close to linear), is shown for several lattice
spacings, two gauge actions, and three renormalized anisotropies. All show
the same result up to tiny cut-off effects. To linear order the
curve can be parametrized as
\begin{equation}
% R_E/\xi_w^2 \approx 1+ 0.296 (\xi_g^2/\xi_w^2-1)\,.
\xi_g/\xi_w = 1+ 1.71( R_E/\xi_w^2 -1)\,.
\label{eq:REparam}
\end{equation}

This confirms our assumption that the universality of $R_E/\xi_w^2$ is not
restricted to the case where $\xi_g=\xi_w$ and thus the flow is isotropic in
physical units.

Especially in full QCD, the Wilson flow analysis is significantly cheaper than
the generation of independent configurations, and it is normally not an
obstacle to calculate $R_E$ at several $\xi_w$ parameters. In some cases,
like quenched QCD, the flow integration may seem expensive.
The determination of $\xi_g$ can then be greatly simplified if the 
relation is known between $R_E$, $\xi_w$ and $\xi_g$.
One then tries to guess $\xi_g$ to, say, 10\% accuracy (the typical size of
radiative corrections to the anisotropy with improved actions), and measures
$R_E$. Figure \ref{fig:universal} can then be used to determine $\xi_g/\xi_w$ and
thus also $\xi_g$.

The universality is not guaranteed between quenched and full QCD.
Nevertheless we find that the effect of the quarks on this particular relation
is mild. To illustrate this we plot a staggered data set 
($a=0.12~fm$, physical quark masses, 2+1 flavors). Though not compatible with
the quenched data, it is remarkably close.

The basic tool in our tuning procedure, the Wilson flow, was also advertised as
``gradient flow'' in Reference \cite{Luscher:2011bx}. This name refers to the fact
that the generator of the stout smearing in Equation (\ref{eq:flow_x}) is the
gradient of the widely used plaquette gauge action: $X=-\partial S_g/\partial
U$.  In Reference \cite{Borsanyi:2012zs} we have already given up on this
correspondence by using the gluonic flow equation in full QCD. We also provided
numerical evidence for the irrelevance of the improvement terms in the gauge
action that is used to construct the flow equation. The flow based on the
derivative of the tree-level Symanzik action gave precisely the same
continuum limit, as the simpler flow based on the plaquette action. This was
checked on configurations generated using the Wilson as well as the
Symanzik gauge action, with our without quarks.

Here we manifestly break the concept of the ``gradient flow'' by using the
renormalized anisotropy in the flow equation, in contrast to the action where
the bare anisotropy is used.  In this presentation the Wilson flow is part of
the observable, that selects a particular macroscopic scale in both time and
spacelike directions. Our method is expected to work independently on the
details of the action that was used to generate the configurations. If we used
$\xi_g^{(0)}$ in the Wilson flow, $R_E$ gave very different anisotropies, that
are quite incompatible with $\xi_K$ in Figure \ref{tab:plaq}. Actually
Figure \ref{fig:universal} can also be used to predict this behaviour:
Substituting $\xi_w$ by $\xi_g^{(0)}$ in the parametrization (\ref{eq:REparam})
and performing a linear approximation in $\xi_g^{(0)}/\xi_g^2-1$ 
one finds $R_E/\xi_g^2 \approx 1+ 0.71\left( (\xi_g^{(0)}/\xi_g)^2-1\right)$.
Using the numbers in Table \ref{tab:plaq} we find that $\sqrt{R_E}$ is about
12\% lower than with the correct definition.

%]]]

\section{Parameter tuning in the quenched case}%[[[

In the pure gluonic theory there is only one extra bare parameter that
is induced by the 3+1-anisotropy. In this Section we tune this bare parameter
such that $\xi_g$ is equal to a predefined target value.
Keeping $\xi_g$ constant over a range of lattice spacings is a particularly
important ingredient of a continuum extrapolated lattice result.

In the course of tuning one could determine $\xi_g$ for several bare anisotropy
parameters ($\xi_0$), and find the preferred choice through interpolation.
The procedure is somewhat simplified in the sense that for every $\xi_0$
the Wilson flow is integrated once only. In fact, we may use the target
anisotropy  $\xi_w=\xi_g$ for all bare anisotropies. The configurations, that
are generated on the fly for the flow integration need not be stored. Again,
$R_E/\xi_g^2$ is measured and interpolated in $\xi_0$. The equation
$R_E=\xi_g^2$ locates the point where $\xi_0$ is accepted.
 
In this Section we tabulate the tuned anisotropies that we determined
for the tree-level Symanzik gauge action. For a limited number of gauge couplings
References \cite{Sakai:2000jm,Sakai:2003va} gives the tuned bare anisotropies
for this action as well as the Iwasasaki and DBW2 actions, though the main
focus there was to establish the perturbative regime. With our method we
can give $\xi_g^{(0)}$ with sub-percent precision, accompanied with the scale
setting. Since the anisotropy is defined on the $w_0$ scale, this lengh
scale has to be resolved by the lattice and contained by the box.
This condition, however, rules out the discussion of perturbative gauge
couplings. 

In our set of simulations the lattice box size was always larger than $8w_0$. The
aspect ratio of the lattice matched the renormalized anisotropy. The quenched
configurations were generated on the QPACE machine with a separation of 50
updates (each consisting of 1 heatbath + 4 overrelaxation sweeps) between
measurements. $\xi_g^{(0)}$ was seeked in four or more points in the range
$\pm 20$\% around the estimated bare anisotropy. We used a quadratic fit
for the interpolation in $\xi_g^{(0)}$.
Our results are given in Table \ref{tab:xi0}.

\begin{sidewaystable}
\begin{center}
\begin{tabular}{|c||c||c|c||c|c||c|c|}
\hline
\multirow{2}{*}{$\beta$}&
$\xi_g=1$&
\multicolumn{2}{|c||}{$\xi_g=2$}&
\multicolumn{2}{|c||}{$\xi_g=3$}&
\multicolumn{2}{|c|}{$\xi_g=4$}\\
\cline{2-8}
&$w_0$
&$w_0$ &$\xi_0$
&$w_0$ &$\xi_0$
&$w_0$ &$\xi_0$\\
\hline
% xi= 1    xi = 2                      xi = 3                       xi = 4
4.20& 1.2607( 3) &1.0000( 6)&1.8070(31)  &0.9340( 1)&2.6415( 7)   &0.8527( 2)&3.4380(28)\\
4.30& 1.4851( 3) &1.1823( 3)&1.8205(14)  &1.0671(10)&2.6513(47)   &1.0153( 5)&3.4851(37)         \\
4.46& 1.8877(14) &1.5199( 6)&1.8293(14)  &1.3814( 6)&2.6714(21)   &1.3188( 6)&3.5221(31)         \\
4.60& 2.3002(22) &1.8612(16)&1.8386(32)  &1.6946( 6)&2.6880(51)   &1.6242(19)&3.5480(70)         \\
4.70& 2.6211(24) &2.1409(27)&1.8461(22)  &1.9540(20)&2.6899(49)   &1.8721(30)&3.5699(109)         \\
4.81&3.0295(34)  &2.4779(49)&1.8457(56)  &2.2708(28)&2.7102(42)   &2.1690(59)&3.5685(122)         \\
4.90&3.4200(98)&2.7945(54)&1.8554(35)  &2.5531(32)&2.7273(94)   &2.4511(48)&3.5799(121)         \\
5.00&3.8830(137)&3.1732(112)&1.8554(92) &2.9163(39)&2.7319(105)  &2.7974(53)&3.5890(108)         \\
5.10&4.3727(242)&3.5886(112)&1.8649(143)&3.3482(67)&2.7461(87)   &3.2122(75)&3.6144(166)         \\
5.20&4.9609(498)& -         & -         &3.8628(200)&2.7460(115) &3.6359(148)&3.6016(302)       \\
\hline
\end{tabular}
\caption{\label{tab:xi0}
The scale and bare anisotropy at various gauge couplings and target
anisotropies of the tree-level Symanzik gauge action. In this action we keep
the (1x2) rectangles in all orinentations, thus with $\xi_0=1$ the isotropy is
completely restored. For comparison and other possible uses we give
the $w_0$ scale on isotropic lattices, as well.
}
\end{center}
\end{sidewaystable}

%]]]

\section{Fermion anisotropy}

If quarks are considered on an anisotropic lattice, then a bare anisotropy
parameter $\xi_f^{(0)}$ has to be introduced in the quark action and tuned as
the lattice spacing is changed. We use a clover improved Wilson quark action
for this study, the definition can be found in Equation (\ref{eq:dirac}). The
anisotropy of the lattice measured from observables, that are built up from the
quarks, is called fermion anisotropy, $\xi_f$.  The bare parameters of the
action have to be tuned such, that the renormalized anisotropies, $\xi_f$ and $\xi_g$
are equal.

\begin{figure}
\begin{center}
\includegraphics[width=7.2cm]{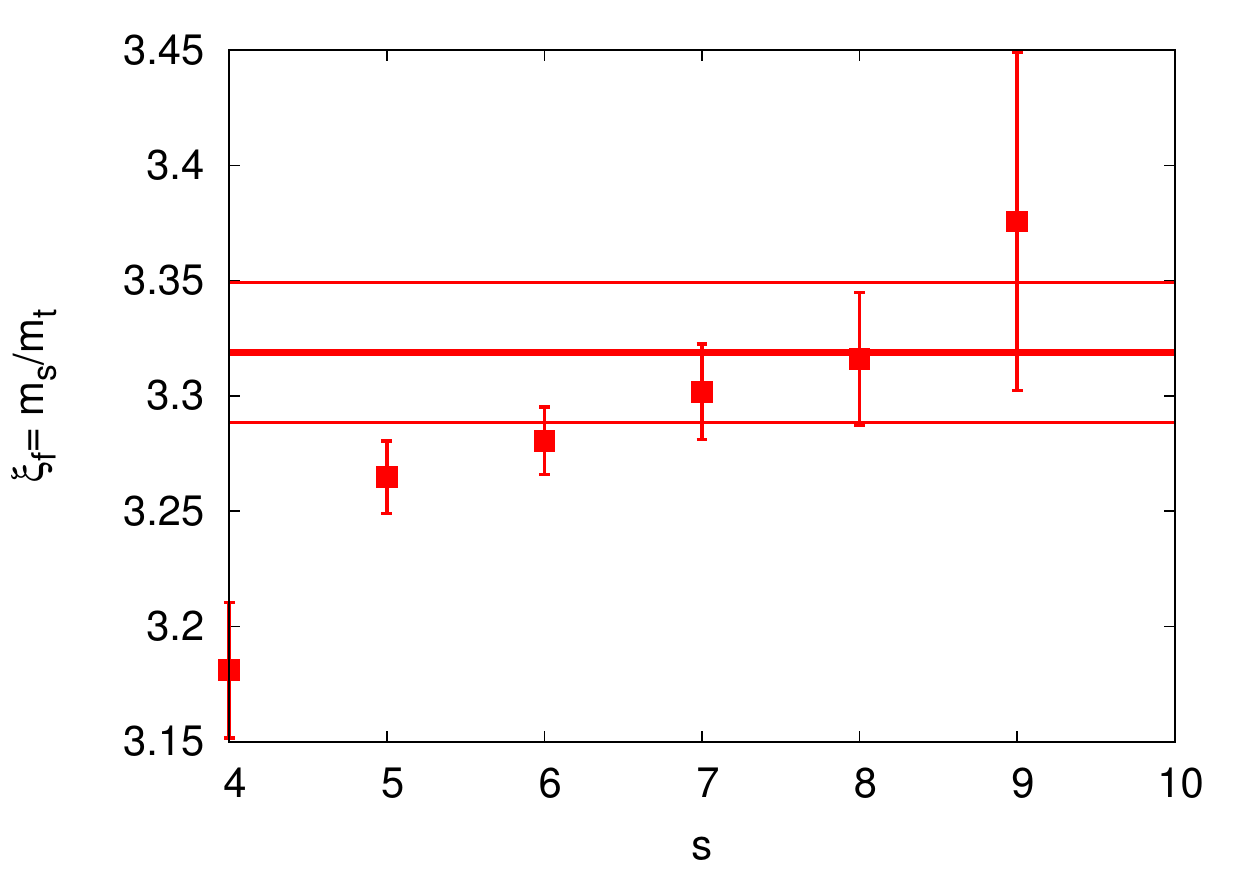}
\includegraphics[width=7.2cm]{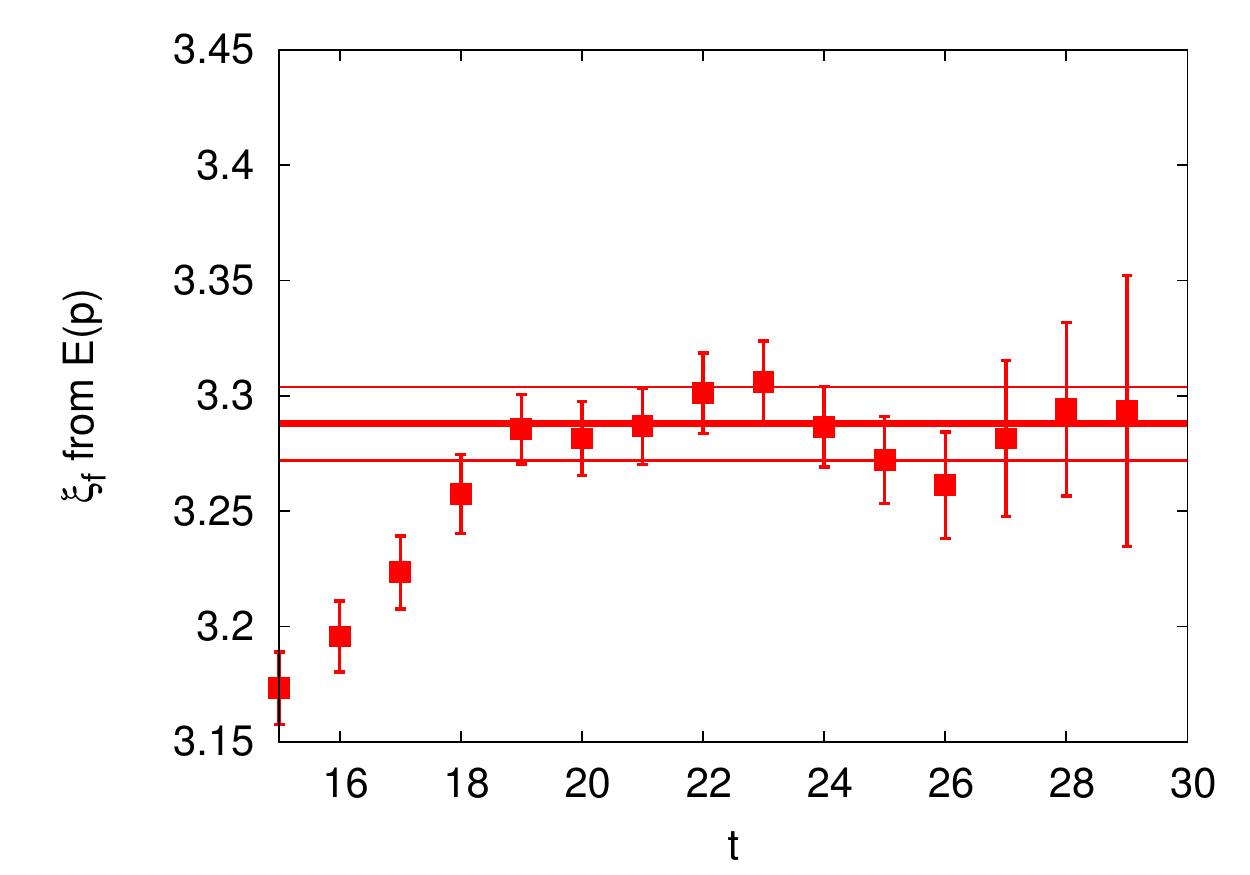}
\caption{
\label{fig:xif}
%\label{fig:msmt} 
Left: Fermion anisotropy determined from Equation (\ref{eq:xif1}).
%\label{fig:clight} 
Right: Fermion anisotropy determined from Equation (\ref{eq:xif2}).
}
\end{center}
\end{figure}

For the fermion anisotropy $\xi_f$ one usually extracts the masses from the
asymptotic decay of a hadron correlator in the spatial and temporal directions:
\bea
\label{eq:xif1}
m_s/m_t= \xi_f,
\eea
where $m_s$ and $m_t$ are the masses in the spatial and temporal directions.
In practice we consider the standard effective masses in both directions,
and for each spatial separation $s$ we solve the equation
\bea
m_s(s)/m_t(s\cdot\xi_f(s))= \xi_f(s),
\eea
for $\xi_f(s)$, which we call ``effective anisotropy''. For the solution the temporal mass
is interpolated to non-integer arguments. The fermion anisotropy is then defined
as the plateau of the effective anisotropy as $s\to\infty$.
Alternatively
one can also measure the hadron energy for nonzero momenta from the temporal 
hadron correlator, and define $\xi_f$ as
\bea
\label{eq:xif2}
E_t^2(p)= m_t^2 + \frac{p^2}{\xi_f^2}.
\eea
This can be done for each separation in time, so one obtains an effective anisotropy plot again. 
The two definitions might differ in lattice artefacts, that are proportional to the lattice spacing.

We illustrate these two methods on 
quenched configurations generated with tree level improved Symanzik gauge
action. The parameters are
\begin{center}
\begin{tabular}{|c|c|}
\hline
lattice size  & $20^3\times60$\\
$\beta$       & 4.46          \\
$\xi_g^{(0)}$ & 2.674         \\
\hline
\end{tabular}
\end{center}
We obtain for $w_0/a_s= 1.379(2)$ and $\xi_g=2.999(8)$. The hadron, we choose
for the $\xi_f$ determination, is the pseudoscalar meson with the mass set
approximately to $m_s=0.25$. The bare fermion anisotropy is set to
$\xi_f^{(0)}=3$. The figures correspond to $0.13$ fm spatial lattice spacing
and $390$ MeV meson mass.  Figure \ref{fig:xif} shows the effective anisotropy
extracted from the ratio of spatial and temporal masses in the left panel and
from the dispersion relation in the right panel.

We now introduce gauge links smearing in the quark operator on the anisotropic
lattice. We hope for the same improvements, as it was the case on isotropic
lattices: smeared link actions have improved chiral properties, renormalization
constants are closer to the tree level values, so as the clover coefficient of
the non-perturbative $O(a)$-improvement. Additionally we expect, that the tuned
bare quark anisotropy parameter $\xi_f^{(0)}$, for which $\xi_f=\xi_g$ holds,
is closer to its tree level value, ie. to $\xi_f$.

\begin{figure}[t]
\centerline{\includegraphics{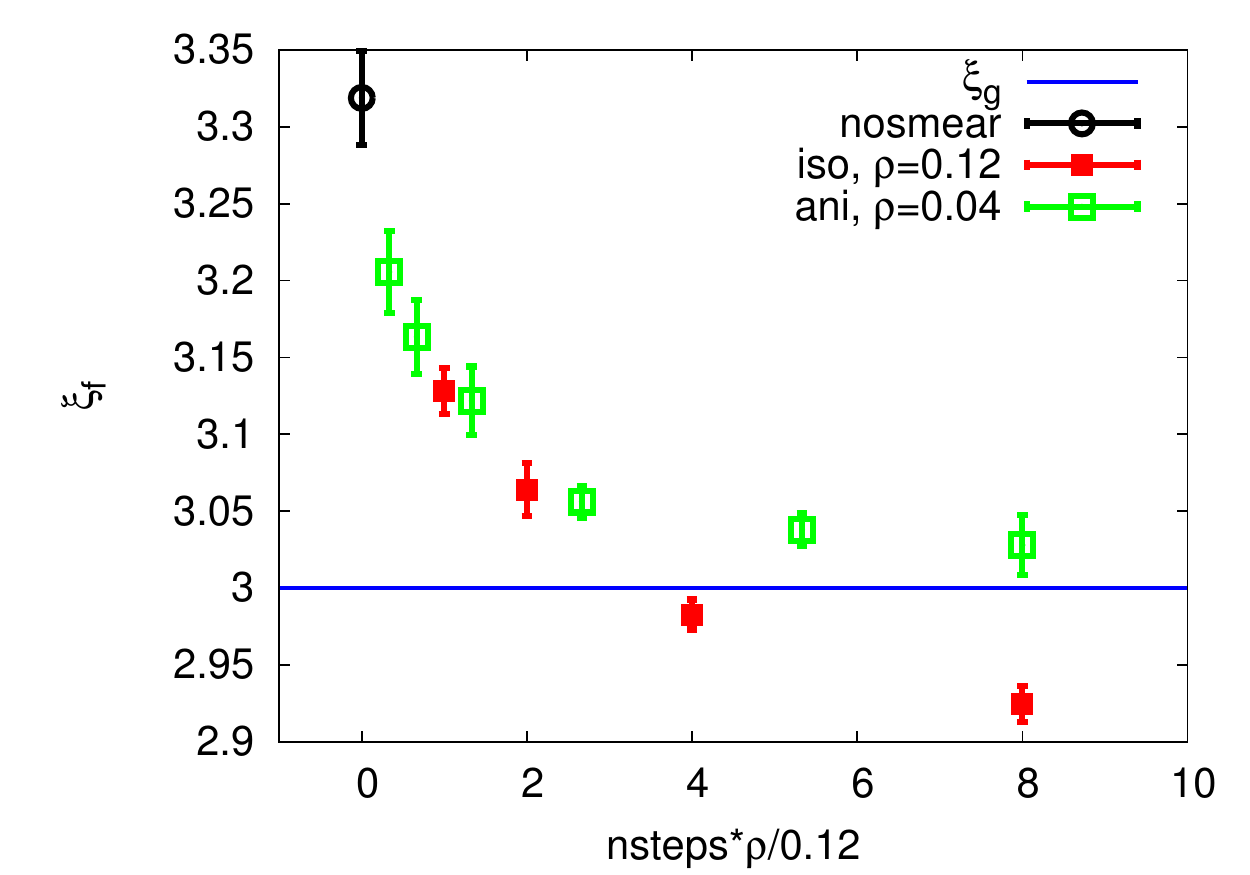}}
\caption{\label{fig:xifsm}
The renormalized fermion anisotropy as the function of smearing steps for ``isotropic''
and ``anisotropic'' stout smearing. In the latter case the number of steps is scaled by a factor 3.
}
\end{figure}

The expectation is confirmed in a numerical experiment.  We fix the bare
fermion anisotropy to the gauge anisotropy, $\xi_f^{(0)}=\xi_g$, and study the
renormalized fermion anisotropy for different number of ``isotropic'' stout smearing
steps with parameters $\rho_{ij}=\rho_{i4}=\rho_{4i}=\rho$. In each
case the pseudoscalar mass was tuned to $m_s=0.25$ again.  Figure
\ref{fig:xifsm} shows, that increasing the number of steps brings the fermion
anisotropy closer to the gauge anisotropy, at four smearing steps their
difference is less than 1\%. This means that using this particular smearing,
the tuning condition $\xi_f=\xi_g$ is satisfied to 1\% precision without tuning
the bare fermion anisotropy.  The effect of changing the pseudoscalar mass by a
factor two upwards is about on the level of the statistical error.

Interestingly as one increases the number of smearing steps beyond four, the
fermion anisotropy decreases further and the tuning of the anisotropy parameter
becomes neccesary again (now in the other direction). ``Isotropic'' stout smearing
washes out the anisotropy of the backgroung gauge configuration.

This encourages us to consider stout smearing, with anisotropic parameters. A
natural choice is to use the generator of the Wilson flow, ie. Equation
(\ref{eq:flow_x}) with coefficients $\rho_{i4}=\xi_g^2\rho$ and
$\rho_{ij}=\rho_{4i}=\rho$, as generator of the stout smearing transformation.
There is however an important limitation. As it is known, stout smearing gets
unstable for large smearing parameters.  In our current numerical study we
found, that the boundary of the instability region for the $\rho$ parameter is
reduced by a factor $3$. In order to keep the strength of the smearing
constant, we are forced to increase the number of smearing steps by the same
factor. On Figure \ref{fig:xifsm} we plot the results with this ``anisotropic''
stout smearing. As it can be seen it also brings the fermion anisotropy closer
to the gauge anisotropy. Differently from the ``isotropic'' smearing, it does
not get worse for larger number of smearing steps, the neccesary tuning is
getting gradually smaller as the number of smearing steps is increased.

Let us emphasize here, that the continuum limit is universal regardless of the
the details of the smearing in the Dirac operator. It is of practical
importance to use a smearing, where the anisotropy renormalization is
suppressed.

\section{Parameter tuning with dynamical quarks}
\begin{figure}
\centerline{\includegraphics{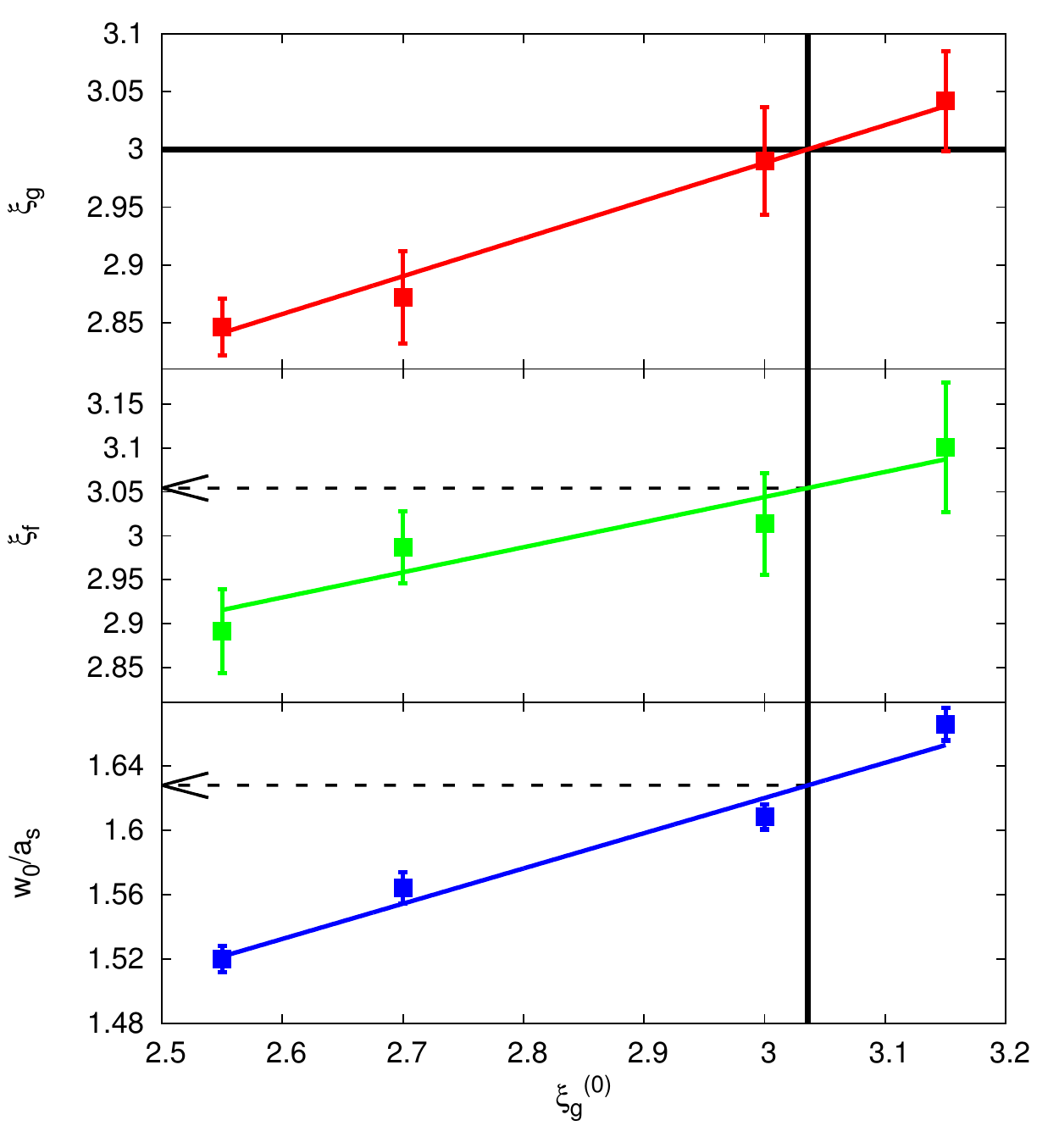}}
\caption{\label{fig:tune}
The gauge anisotropy, the fermion anisotropy and the $w_0$-scale as the function
of the bare gauge anisotropy in our runs with dynamical quarks.
}
\end{figure}

In this section we propose and test a strategy to tune the anisotropy
parameters for $n_f=3$ degenerate flavors of dynamical Wilson quarks.

As we have seen in the previous section, up to some precision there is no need
to tune the bare fermion anisotropy, if the gauge link smearing in the Dirac
operator is properly chosen. This can be achieved either by using ``anisotropic''
smearing with a high enough number of smearing steps. Or one can use ``isotropic''
smearing, it also reduces the necessary tuning until some number of smearing
steps. In the latter case one has to be careful no to overdo the smearing. The bare
fermion anisotropy $\xi_f^{(0)}$ is to be set to the target anisotropy.  Only
one anisotropy parameter, the bare gauge anisotropy $\xi_g^{(0)}$ has to be
tuned until the renormalized gauge anisotropy $\xi_g$ equals the target
anisotropy.

We choose to use ``isotropic'' stout smearing in the Dirac opeartor to test the above
strategy and take the same number of smearing steps, which has turned out to be
optimal in the quenched case (ie. four steps with $\rho=0.12$). We generated
ensembles with tree level improved Symanzik gauge action and $n_f=3$
flavors of dynamical Wilson quarks.  The target anisotropy is $3$, so we set
$\xi_f^{(0)}=3$. We use Rational Hybrid Monte Carlo
algorithm combined with Hasenbusch preconditioning for the generation of
configurations. The parameters which we use for the tuning are
\begin{center}
\begin{tabular}{|c|c|}
\hline
lattice size  & $16^3\times96$ \\
$\beta$       & 3.5 \\
$m$           & -0.025 \\
$\xi_f^{(0)}$ & 3.00 \\
$\xi_g^{(0)}$ & 2.55, 2.70, 3.00, 3.15 \\
\hline
\end{tabular}
\end{center}

On the upper panel of Figure \ref{fig:tune} we plot the renormalized gauge
anisotropy as function of the bare anisotropy in our four runs. The results can
be interpolated by a linear fit. The gauge anisotropy takes the target value
$\xi_g=3$, when the bare anisotropy is set to
$\xi_g^{(0)}=3.04(x)$\footnote{Suprisingly the bare and renormalized gauge
anisotropies are consistent. It is most probably an accident, but it deserves
more investigation.}. At this parameter the fermion anisotropy is
$\xi_f=3.05(3)$, which is less than 2\% and somewhat more than 1$\sigma$ away
from the desired point (where $\xi_f=\xi_g$). The $w_0$-scale at this point is
$w_0=1.63(3)$, this corresponds to a spatial lattice spacing of $0.11$ fm. The
pseudoscalar mass is $485$ MeV.

We conclude that gauge link smearing also helps to decrease the fermion
anisotropy renormalization in the dynamical case. In our concrete case no
tuning of $\xi_f^{(0)}$ is needed, if the required precision is not better than
$2\%$.

\section{Conclusions}

In this paper we generalized the $w_0$ scale to anisotropic lattices. We
found that the scale setting procedure requires the knowledge of the
anisotropy. We worked out a method to extract the renormalized anisotropy
form the Wilson flow, which is also the basis of our scale setting.
Our method is in agreement with the standard results based on ratios
of Wilson loops, but does not rely on the interpolation of lattice data
between lattice sites, nor does it require to evaluate large Wilson loops.

To illustrate the use of our measure for gauge anisotropy we tabulated
the bare anisotropies of the tree-level Symanzik gauge action 
for three renormalized anisotropies. We also determined the scale setting
$w_0$ and thus made the parameters available for immediate use.

We have studied the effect of smearing on the fermion anisotropy. We have
observed on quenched configurations that both ``isotropic'' and ``anisotropic''
smearing significantly reduces the difference between the bare and renormalized
anisotropy. This statement was also verified with three flavor dynamical
simulations. In the future we will extend this study by determining the bare
anisotropies with dynamical fermions for a wider range of lattice spacings and
renormalized asymmetries.

\appendix
\section{Lattice actions}
The anisotropic
gauge action is
\bea
\frac{\beta}{\xi_g^{(0)}} \sum_{x,i<j} \left[ 1 - \frac{1}{3}{\rm Re tr} {\mathcal U}_{ij}(x)\right] 
+\beta \xi_g^{(0)} \sum_{x,i} \left[ 1 - \frac{1}{3}{\rm Re tr}{\mathcal U}_{i4}(x)\right],
\eea
where the $\mathcal{U}_{\mu\nu}(x)$ loop operator is constructed from gauge links along plaquettes and rectangles:
\bea
\mathcal{U}_{\mu\nu}= c_0 W_{\mu\nu}(1,1) + c_1 W_{\mu\nu}(1,2) + c_1 W_{\mu\nu}(2,1).
\eea
We use both the simple plaquette action $c_0= 1, c_1=0$ and tree level Symanzik improved action $c_0=5/3, c_1=-1/12$ in this paper.

We choose the anisotropic Wilson-clover Dirac operator as
\bea
(D \psi)_x &=& 
(m+3+\xi_f^{(0)})\psi_x -\nonumber\\
&-&\frac{\xi_f^{(0)}}{2}  \left[ (1+\gamma_4)U_{4}(x)\psi_{x+i} + (1-\gamma_4)U^\dagger_{4}(x-i)\psi_{x-i}\right] -\nonumber\\
&-&\frac{1}{2}\sum_i \left[ (1+\gamma_i)U_{i}(x)\psi_{x+i} + (1-\gamma_i)U^\dagger_{i}(x-i)\psi_{x-i}\right] -\nonumber\\
&-&\frac{1}{2} \sum_{i>j} c^{SW}_{s} f_{ij}(x) \sigma_{ij} \psi_x -\frac{1}{2} \sum_{i} c^{SW}_{t} f_{4i}(x) \sigma_{4i} \psi_x,
\label{eq:dirac}
\eea
where the Dirac-sigma matrices are $\sigma_{\mu\nu}=1/2
[\gamma_\mu,\gamma_\nu]$ and the $f_{\mu\nu}(x)$ loop operator is the
discretization of the field-strength tensor built up from gauge links along the
clover path. For the clover coefficients we choose $c^{SW}_{s}=1$ and
$c^{SW}_{t}=(\xi_f^{(0)}+1)/2$. Our definition is very similar to that of Reference
\cite{Edwards:2008ja}, the difference is, that they use $c^{SW}_{t}= (\xi_g^{(0)}/\xi_f^{(0)} + \xi^*)/2$, where $\xi^*$ is
the target anisotropy.

\section*{Acknowledgments}

We thank Christian Hoelbling for useful discussions. This research has been
partly supported by the European Research Council grant 208740 (FP7/2007-2013).
Computations were performed on the Blue Gene supercomputers at FZ J\"ulich on
the QPACE facility supported by the Sonderforschungsbereich TR55 and on GPU
\cite{Egri:2006zm} clusters at the Wuppertal University.

\bibliographystyle{JHEP}
\bibliography{notears}

\end{document}